\documentclass{article}
\usepackage{spconf,amsmath,graphicx}
\usepackage{cite}
\usepackage{url}
\usepackage{graphicx}
\usepackage{amssymb,amsmath,bm}
\usepackage{textcomp}
\usepackage{booktabs}
\usepackage{hyperref}
\usepackage{verbatim}
\usepackage{multirow}
\usepackage[table,xcdraw]{xcolor}
\usepackage{array}
\usepackage{tabularx}
\usepackage{bbding}
\usepackage{xspace}
\usepackage{color} 
\newcolumntype{Y}{>{\centering\arraybackslash}X}
\newcommand{\norm}[1]{\left\lVert#1\right\rVert}
\makeatletter

\DeclareRobustCommand\onedot{\futurelet\@let@token\@onedot}
\def\@onedot{\ifx\@let@token.\else.\null\fi\xspace}
 
\def\ie{\emph{i.e}\onedot}

\makeatother

\title{ScoreDec: A Phase-preserving High-Fidelity Audio Codec with A Generalized Score-based Diffusion Post-filter}
\name{Yi-Chiao Wu, Dejan Markovi\'{c}, Steven Krenn, Israel D. Gebru, Alexander Richard}
\address{Codec Avatars Lab, Pittsburgh PA, USA}
\begin{document}
\ninept
\maketitle
\thispagestyle{plain}
\pagestyle{plain}
\begin{abstract}
Although recent mainstream waveform-domain end-to-end (E2E) neural audio codecs achieve impressive coded audio quality with a very low bitrate, the quality gap between the coded and natural audio is still significant. A generative adversarial network (GAN) training is usually required for these E2E neural codecs because of the difficulty of direct phase modeling. However, such adversarial learning hinders these codecs from preserving the original phase information. To achieve human-level naturalness with a reasonable bitrate, preserve the original phase, and get rid of the tricky and opaque GAN training, we develop a score-based diffusion post-filter (SPF) in the complex spectral domain and combine our previous AudioDec with the SPF to propose ScoreDec, which can be trained using only spectral and score-matching losses. Both the objective and subjective experimental results show that ScoreDec with a 24~kbps bitrate encodes and decodes full-band 48~kHz speech with human-level naturalness and well-preserved phase information.
\end{abstract}
\begin{keywords}
audio codec, phase-preserving codec, codec post-filter, score-based generative model, human-level naturalness
\end{keywords}
\section{Introduction}
\label{sec:intro}
\vspace{-\baselineskip}
Audio signals have a very high temporal resolution. For instance, the standard sample rate of CD stereo 16-bit audio is 44.1~kHz, requiring a 1,411~kbps bitrate for transmission and 172~KB for storing one second of audio. Audio codecs take advantage of the redundancies caused by the quasi-periodic nature of audio to produce low-bitrate and compact codes for efficient audio storage and transmission, and a decoder is needed to convert these codes back into the waveform.

The development of audio codecs dates back decades. In the early days, due to limited transmission bandwidth, most codec designs focused on low-bitrate approaches, trading off audio quality for compact codes. However, with the ever-increasing use of voice and videophone applications, as well as the distribution of audio-visual data for internet video and broadcast applications, several audio codecs and standards have been proposed with an emphasis on better audio quality. Lossless audio codecs~\cite{mpeg4, flac} with compression ratios as low as $2\times$ and lossy audio codecs~\cite{opus, amrwb, evs} with a compression ratio of up to $10\times$ have been proposed based on carefully engineered design choices and handcrafted signal processing components. Although these lossy audio codecs meet the compression requirements of most current applications, the ad~hoc designs and limited modeling capacity of these lossy codecs still result in a significant quality gap between natural and reconstructed audio signals.

To avoid ad~hoc designs and take advantage of the powerful modeling capacity of neural networks (NNs), end-to-end (E2E) neural audio codecs~\cite{aecodec2018, vqvae2019, cmrl, soundstream, encodec, audiodec} in the waveform domain recently have been intensively investigated. Although these neural codecs achieve impressive coded audio quality in very low bitrate conditions (e.g., 3--8~kbps), there are three main problems, \ie saturation in quality, tractability, and phase preservation. Most neural codecs control the bitrate by adopting different numbers of codebooks~\cite{rvq} while keeping the same temporal resolution of the codes because of the fixed network architecture. As a result, these neural codecs tend to underperform the digital-signal-processing (DSP)-based codecs and quickly reach a quality saturation point when the bitrate is gradually increased to the operation bitrate of the DSP-based codecs (e.g., 24~kbps for Mono Opus~\cite{opus}), which is a reasonable bitrate for most current systems.  Additionally, the neural codecs usually rely on a generative adversarial network (GAN)~\cite{gan} training to achieve high-fidelity audio reconstruction. However, because of the indirect objective function of fooling the discriminators and the lack of explicit phase modeling, the generators tend to generate a plausible phase instead of the original phase. Since the sound directivity and spatiality reconstructions highly depend on the multi-channel phase information, the broken phase relationships result in significant modeling errors in binaural audio and ambient sound field codings.

Based on the success of score-based diffusion generative models (SGMs)~\cite{ddpm, sgm}, many score-based speech generative~\cite{diffwave, wavegrad} and enhancement~\cite{ sgmse1, sgmse2, cdse} models have been proposed. Notely, the score-based generative model for speech enhancement (SGMSE)~\cite{sgmse1, sgmse2 } achieves very impressive performance for restoring the original phase by explicitly tackling complex spectral restorations. To take advantage of the precise phase modeling of SGMSE, we develop a score-based diffusion post-filter (SPF) in the complex spectral domain for the E2E AudioDec codec~\cite{audiodec}. The proposed ScoreDec attains high-fidelity speech reconstruction, preserves the original phase information, and gets rid of the tricky GAN training.

According to the objective and subjective experimental results, the reconstructed coded speech achieves human-level naturalness with a significantly lower waveform difference from the input natural speech. The effectiveness of the proposed SPF for the DSP-based Opus codec~\cite{opus} is also evaluated to demonstrate its generality. The main contributions of this paper are as follows: (1) We propose a score-based diffusion post-filter that significantly improves the quality of both neural~\cite{audiodec} and DSP-based~\cite{opus} audio codecs; (2) The whole system can be trained with only metric losses in an interpretable manner and the tricky adversarial training is not required; (3) ScoreDec well preserves phase information, resulting in highly accurate original waveform reconstruction.

\section{Background}
\label{sec:backgroud}
\vspace{-\baselineskip}
In this section, the neural audio codec and score-based generative model backbones of the proposed ScoreDec are briefly introduced.

\begin{figure}[t]
\centering
\centerline{\includegraphics[width=0.95\columnwidth]{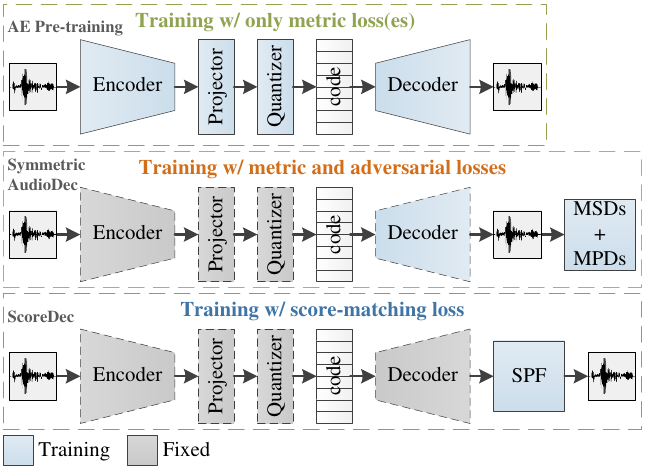}}
\caption{Comparison between symmetric AudioDec and ScoreDec.}
\label{fig:scoredec}
\end{figure}

\subsection{AudioDec}
As shown in Fig.~\ref{fig:scoredec}, AudioDec~\cite{audiodec} is a classical neural codec composed of an encoder, quantizer, and decoder, which are trained in an E2E manner in the waveform domain. The main differences between AudioDec and other E2E neural codecs are the adopted efficient training paradigm and the modularized architecture. Specifically, the essential GAN training is the most computation-consuming process, but the GAN training affects mostly the decoder in fine-tuning the waveform details. To improve the training efficiency, AudioDec adopts a two-stage training. In the first stage, the whole autoencoder is trained using only metric losses such as a mel loss to make training fast. In the second stage, only the decoder and the multi-scale and multi-period discriminators (MSDs~\cite{melgan} and MPDs~\cite{hifigan}) are trained while the encoder is fixed. The modularized architecture also enables exchanging the lightweight AudioDec decoder with a powerful vocoder such as HiFiGAN~\cite{hifigan} with only decoder-sided finetuning. In this paper, symmetric AudioDec denotes the model with a symmetric encoder-decoder architecture while AudioDec denotes the encoder-vocoder version as proposed in the original paper~\cite{audiodec}.

\subsection{Score-based Generative Model for Speech Enhancement}
The SGMSE~\cite{sgmse1, sgmse2} includes a forward process to transfer the unknown clean speech distribution into a simple normal distribution and a reverse process to generate the estimated clean speech from sampling the tractable distribution. Specifically, given a clean speech $\boldsymbol{x_0}$ as the initial state, the corresponding noisy speech $\boldsymbol{y}$, a diffusion time step $t\in[0, T]$, and a standard Wiener process \textbf{w}, the stochastic forward process of the SGMSE is defined by an Ornstein-Uhlenbeck variance exploding (OUVE) stochastic differential equation (SDE)~\cite{ouve} as
\begin{align}
\mathrm{d}\boldsymbol{x}_t=
\underbrace{\gamma(\boldsymbol{y}-\boldsymbol{x}_t)}_{:=f(\boldsymbol{x}_t,\boldsymbol{y})}{\mathrm{d}}t
+
\underbrace{\begin{bmatrix}\sigma_{\text{min}}(\frac{\sigma_{\text{max}}}{\sigma_{\text{min}}})^t\sqrt{2\log(\frac{\sigma_{\text{max}}}{\sigma_{\text{min}}})}\end{bmatrix}}_{:=g(t)}\mathrm{d}\textbf{w},
\label{eq:sde}
\end{align}
where $f(\boldsymbol{x}_t,\boldsymbol{y})$ is the drift function, $g(t)$ is the diffusion coefficient, and $\gamma$ and $(\sigma_{\text{min}}, \sigma_{\text{max}})$ are the constant hyperparameters respectively controlling the stiffness and the injected amount of Gaussian noise of the process at each timestep. Furthermore, the corresponding reverse SDE~\cite{rdem, sgm} is formulated as
\begin{align}
\mathrm{d}\boldsymbol{x}_t=
\begin{bmatrix}-f(\boldsymbol{x}_t,\boldsymbol{y})+g(t)^2\nabla_{\boldsymbol{x}_t}\log p_t(\boldsymbol{x}_t)\end{bmatrix}{\mathrm{d}}t
+
g(t)\mathrm{d}{\Bar{\textbf{w}}}.
\label{eq:rsde}
\end{align}
The gradient of the logarithm distribution of $\boldsymbol{x}_t$, $\nabla_{\boldsymbol{x}_t}\log p_t(\boldsymbol{x}_t)$, is called the score function, and ${\Bar{\textbf{w}}}$ is the time-reversed Wiener process.

Because Eq.~\ref{eq:sde} is a Gaussian process, the distribution of state $\boldsymbol{x_t}$  is a normal distribution whose mean has a closed-form solution as
\begin{align}
\mu(\boldsymbol{x}_0, \boldsymbol{y}, t)=e^{-\gamma t}\boldsymbol{x}_0+(1-e^{-\gamma t})\boldsymbol{y},
\label{eq:cmean}
\end{align}
and the variance also has a closed-form solution as
\begin{align}
\sigma(t)^2=\frac{\sigma_{\text{min}}^2\left((\frac{\sigma_{\text{max}}}{\sigma_{\text{min}}})^{2t}-e^{-2\gamma t}\right)\log(\frac{\sigma_{\text{max}}}{\sigma_{\text{min}}})}{\gamma+\log(\frac{\sigma_{\text{max}}}{\sigma_{\text{min}}})}.
\label{eq:cvar}
\end{align}
As a result, the corresponding score function can be formulated as
\begin{align}
\nabla_{\boldsymbol{x}_t}\log p_t(\boldsymbol{x}_t|\boldsymbol{x}_0,\boldsymbol{y})=
-\frac{\boldsymbol{x}_t-\mu(\boldsymbol{x}_0, \boldsymbol{y}, t)}{\sigma(t)^2}.
\label{eq:score}
\end{align}
Although an arbitrary $\boldsymbol{x}_t$ is available based on the given clean-noisy pair $(\boldsymbol{x}_0,\boldsymbol{y})$ in the training stage, the clean speech $\boldsymbol{x}_0$ is agnostic in the inference stage. To estimate the clean speech based on the noisy speech using the reverse process, SGMSE adopts a neural network as the score estimator during the inference. Specifically, given a sampled Gaussian noise $\boldsymbol{z}$, the $\boldsymbol{x}_t$ can be computed by
\begin{align}
\boldsymbol{x}_t=
\mu(\boldsymbol{x}_0, \boldsymbol{y}, t)+\sigma(t)\boldsymbol{z}.
\label{eq:xt}
\end{align}
By substituting Eq.~\ref{eq:xt} into Eq.~\ref{eq:score}, the score estimator $\boldsymbol{s}_\theta$ can be trained by the score-matching~\cite{score_match} objective function
\begin{align}
\mathop{\arg\min}_{\theta}\mathbb{E}_{\boldsymbol{x}_t|(\boldsymbol{x}_0,\boldsymbol{y}),\boldsymbol{y},\boldsymbol{z},\boldsymbol{t}}
\left[\norm{\boldsymbol{s}_\theta(\boldsymbol{x}_t, \boldsymbol{y}, t)+\frac{\boldsymbol{z}}{\sigma(t)}}_{2}^2\right].
\label{eq:score_match}
\end{align}
The central difference between SGMSE and other conditional SGMs~\cite{diffwave, wavegrad, cdse} is the direct incorporation of the speech generative task into the forward and reverse processes.

\begin{figure*}[t]
\fontsize{9pt}{9pt}
\selectfont
{%
\begin{tabularx}{2.0\columnwidth}{@{}p{0.1cm}XXXX@{}}
\rotatebox[origin=c]{90}{{\centering\arraybackslash}Magnitude} 
&  \includegraphics[width=0.5\columnwidth]{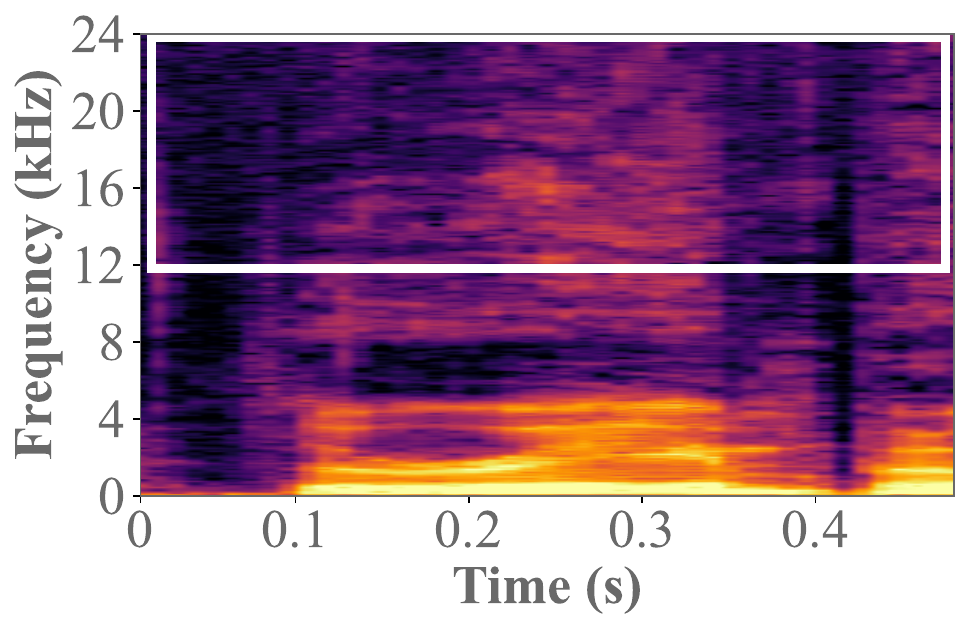}
&  \includegraphics[width=0.5\columnwidth]{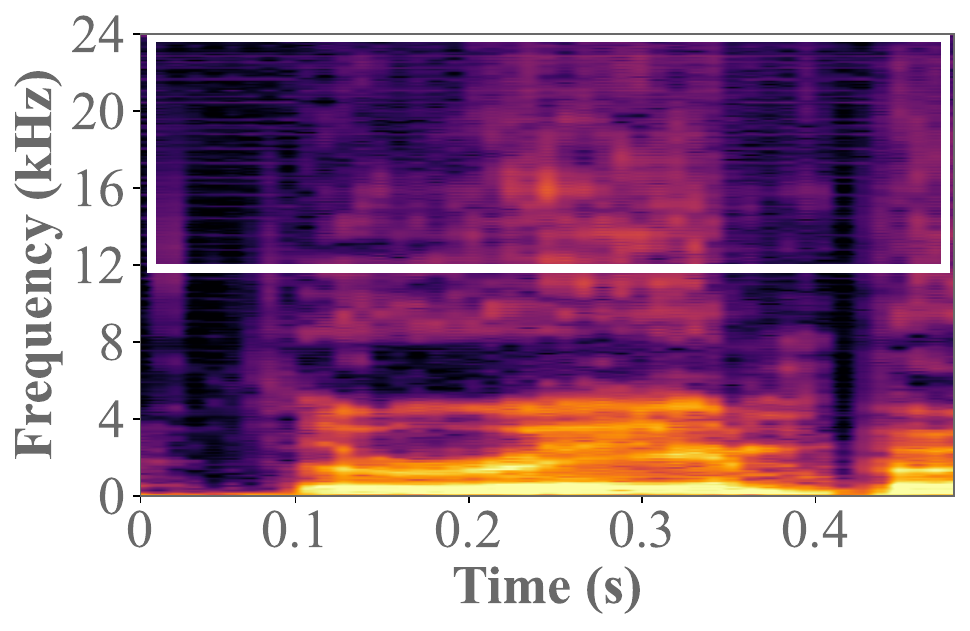}
&  \includegraphics[width=0.5\columnwidth]{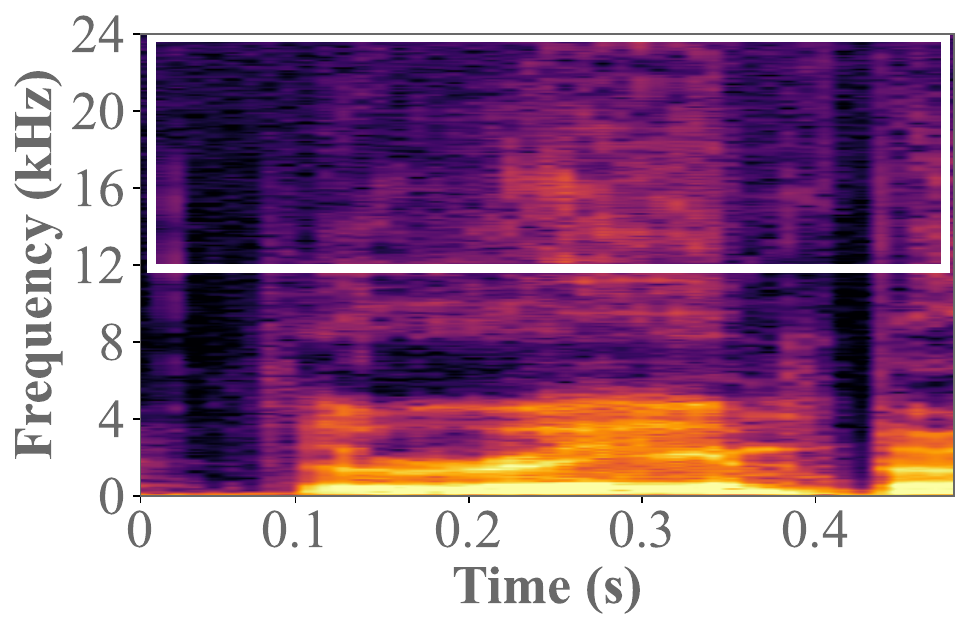}
&  \includegraphics[width=0.5\columnwidth]{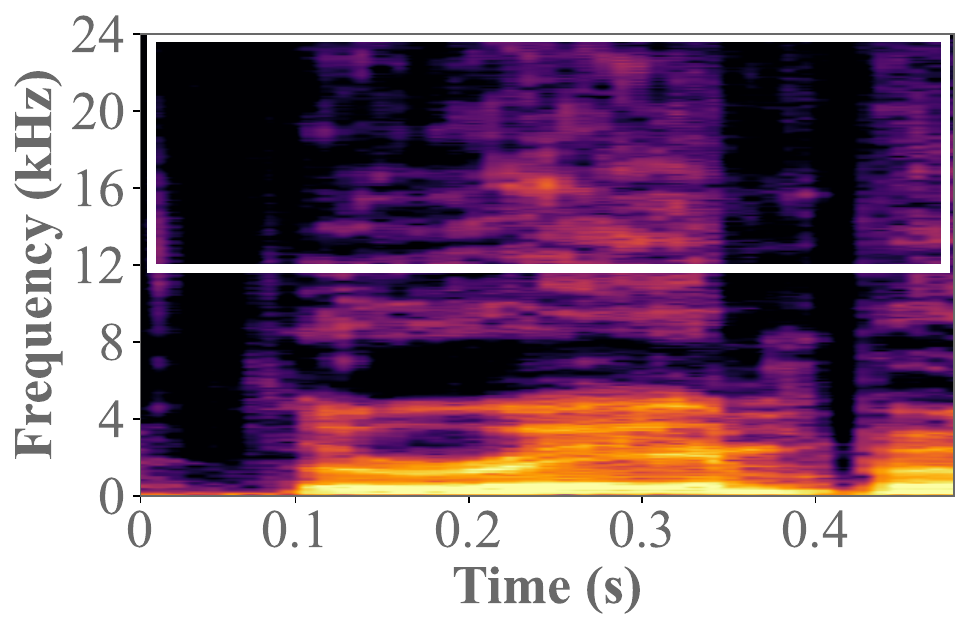}
\\ 

\rotatebox[origin=c]{90}{Phase}
& \includegraphics[width=0.5\columnwidth]{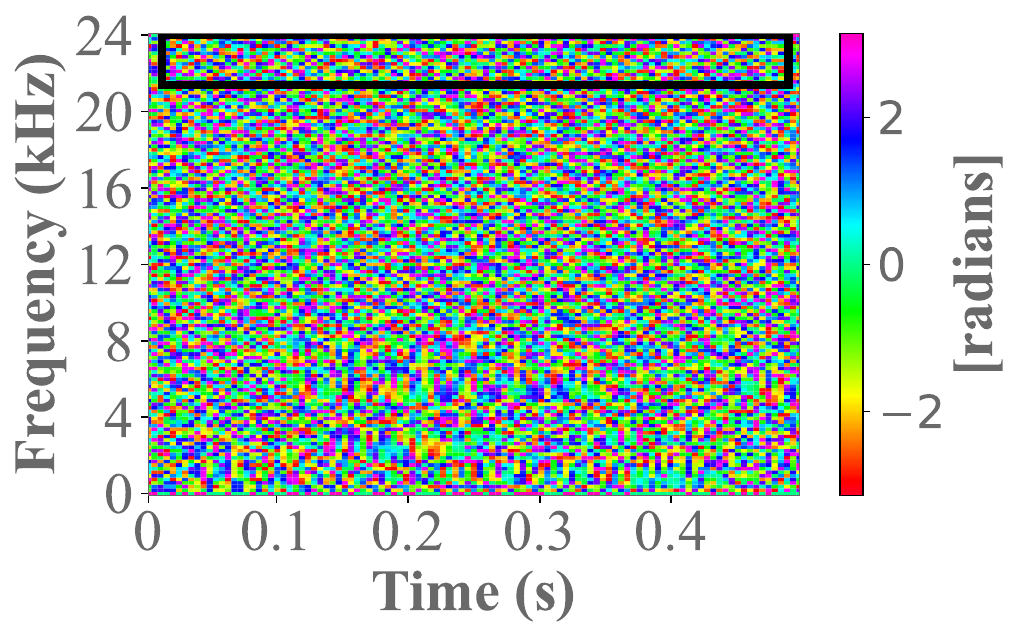}
& \includegraphics[width=0.5\columnwidth]{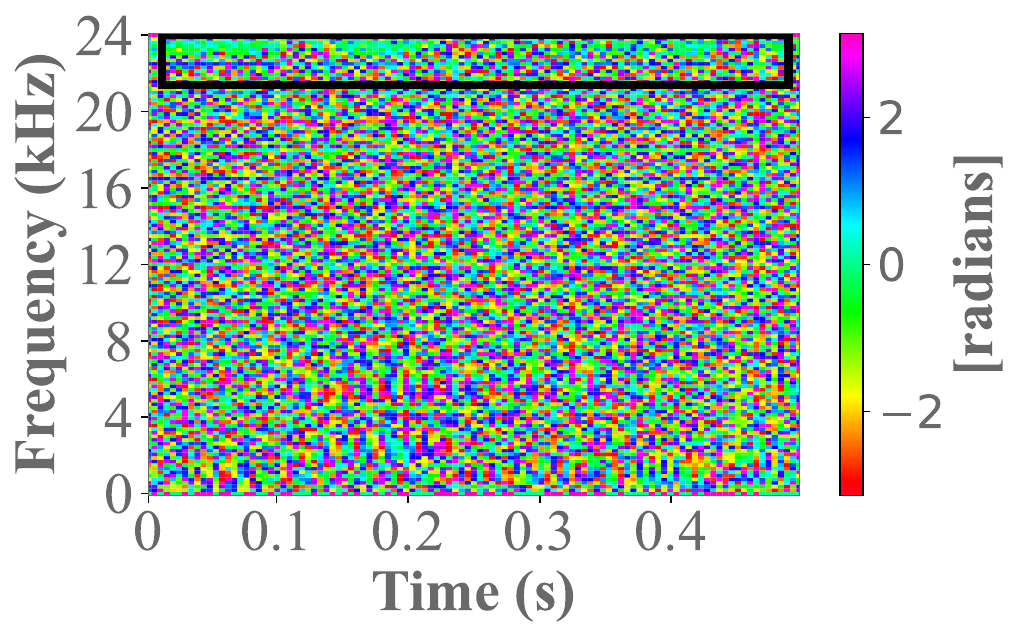} 
& \includegraphics[width=0.5\columnwidth]{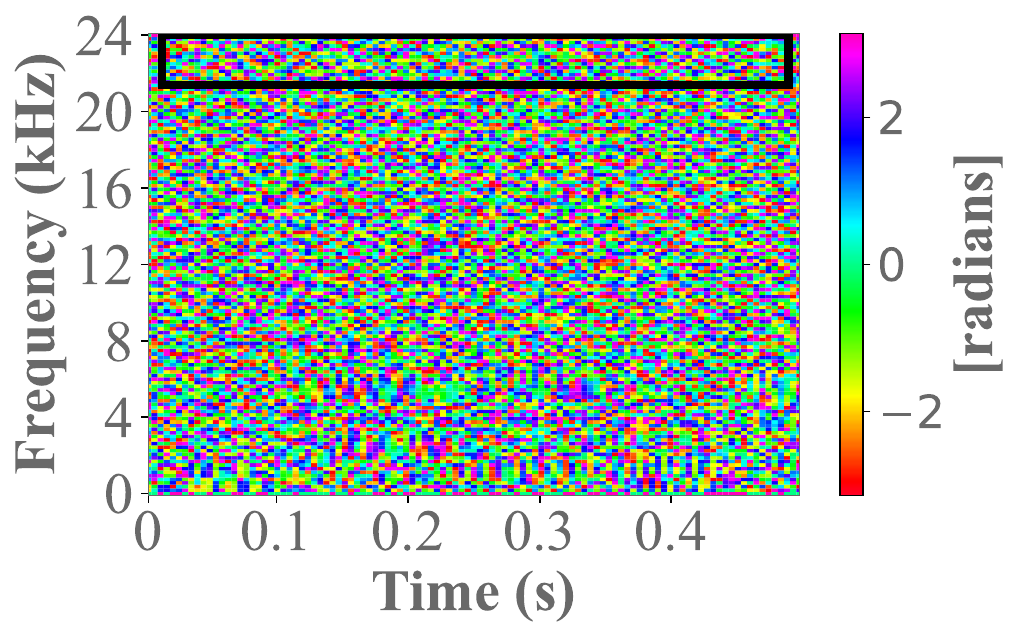}
& \includegraphics[width=0.5\columnwidth]{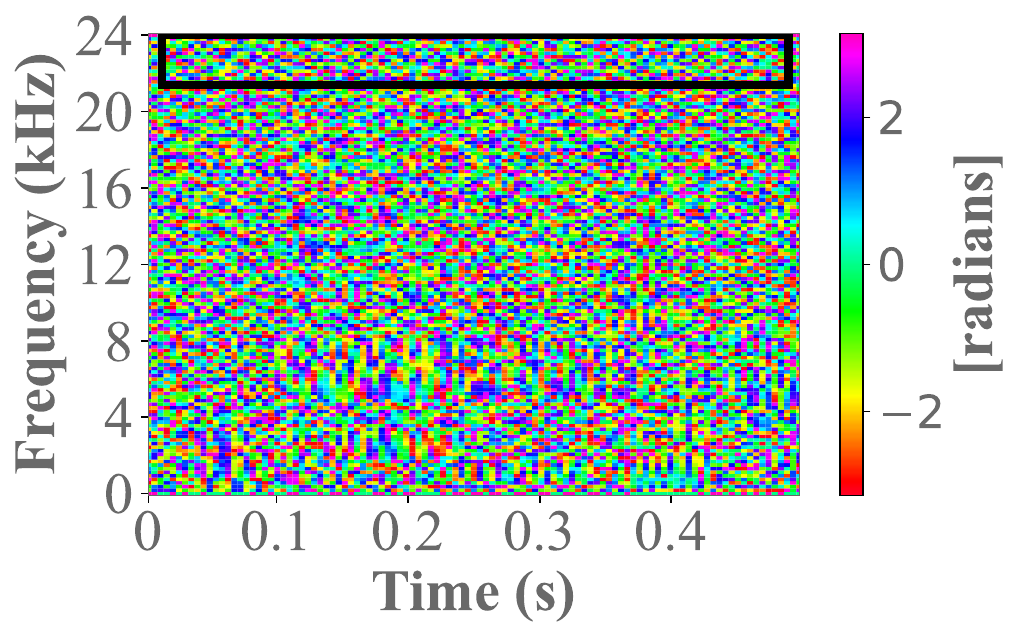}
\\

&  \centering\arraybackslash{(a) Ground truth}
&  \centering\arraybackslash{(b) AudioDec pretraining (only metric loss)}
&  \centering\arraybackslash{(c) fully trained AudioDec (with adversarial loss)}
&  \centering\arraybackslash{(d) ScoreDec}
\\ 
\end{tabularx}%
}
\caption{Comparison of magnitude and phase spectra. (a) Ground truth signal. (b) Metric losses alone over-smoothing frequency spectra and introduce phase artifacts. (c) GAN training compensates for these artifacts but can not faithfully reconstruct details of the original signal. (d) ScoreDec reconstructs subtle high-frequency details and the original phase.}
\label{fig:mag_phase}
\end{figure*}

\section{Method}
\label{sec:proposed}
\vspace{-\baselineskip}
In this section, we first try to answer three related questions: Why is explicit phase modeling difficult? Why is GAN training essential for the current E2E neural codecs? And why is GAN training not preferred? Subsequently, we define the problem tackled in this paper and introduce the proposed method.

\subsection{Problems of GAN-based E2E Neural Codec}
The difficulties of explicit phase modeling mainly root in phase being a white-noise-like signal with a principal value interval $(-\pi, \pi]$. The lack of significant patterns makes phase modeling difficult, and the phase prediction error in angular space should be formulated in a complicated form to account for phase wrapping~\cite{nspp},
\begin{align}
\min\lbrace|\boldsymbol{\hat{p}}-\boldsymbol{p}|,2\pi-|\boldsymbol{\hat{p}}-\boldsymbol{p}|\rbrace.
\label{eq:phase_wrap}
\end{align}
As a result, most current neural codecs adopt only magnitude spectral losses~\cite{audiodec} or losses implicitly tackling the phase such as multi-resolution spectral~\cite{soundstream} and waveform losses~\cite{encodec} and let the GAN training teach the model to learn the magnitude and phase consistency. Since these losses focus on mostly the signal envelope, the model trained with only these losses suffers from over-smoothing and high-frequency aliasing problems. As Fig.~\ref{fig:mag_phase}-(b) illustrates, the AudioDec trained with only the mel spectral loss misses subtle details and generates some undesired details such as high-frequency carrier waves resulting in buzzy sounds. This answers our second question: GAN training is required since these defects can be easily detected by the discriminators, ensuring that the model still generates high-fidelity speech with a reasonable phase (see Fig.~\ref{fig:mag_phase}-(c)). The reason such training is not preferable (question three) is that there is no guarantee of preserving the original phase because of the fuzzy training objective (i.e., maximizing the probability that the discriminators classify the generated speech as natural speech) of the model, and the obscurity of the tricky GAN training hinder the model’s capacity to faithfully reconstruct the input speech, particularly subtle frequency details and original phase.

\subsection{Architecture Overview}
The comparison of the original and predicted spectra in Fig.~\ref{fig:mag_phase} shows that the generated speech of AudioDec without the GAN training (AD\textsubscript{stage1}) includes an over-smoothing and high-frequency-noise corrupted magnitude spectrum and a distorted phase spectrum, which is similar to a classical speech enhancement problem. Therefore, adopting SGMSE as a post-filter to enhance both the real and imaginary spectra, which have clear patterns, of the AD\textsubscript{stage1} coded speech is a reasonable solution to avoid the tricky GAN training and challenging direct phase modeling while restoring the original phase.

The proposed ScoreDec is composed of the symmetric neural audio codec AudioDec~\cite{audiodec} and a score-based diffusion post filter (SPF), which are separately trained using the mel-loss and score-matching loss, respectively. The symmetric AudioDec encodes speech into low-bitrate discrete codes and decodes the preliminary speech for the following SPF processing to generate the final speech. Specifically, a symmetric AudioDec is first trained using only the mel loss (i.e., the 1\textsuperscript{st} stage of AudioDec) as shown in Fig.~\ref{fig:scoredec}. Then, given the paired input waveform $\boldsymbol{x}_{w}$ as the clean speech and the AD\textsubscript{stage1} coded waveform $\boldsymbol{\hat{x}}_{w}$ as the noisy speech, the proposed SPF can be trained using the score-matching objective from Eq.~\ref{eq:score_match}. To train the SPF in the complex spectral domain, the waveform signals $\boldsymbol{x}_{w}$ and $\boldsymbol{\hat{x}}_{w}$ are first transformed into complex spectra $\boldsymbol{x}_{c}$ and $\boldsymbol{\hat{x}}_{c}$ using a short-time Fourier transform (STFT). Following the data representation in~\cite{sgmse1, sgmse2} to avoid the model being dominated by only the high-energy components, an amplitude modulation  
\begin{align}
\boldsymbol{x}_{a}=\beta|\boldsymbol{x}_c|^\alpha e^{i\angle(\boldsymbol{x}_c)}
\label{eq:am}
\end{align}
is applied to both $\boldsymbol{x}_{c}$ and $\boldsymbol{\hat{x}}_{c}$, where $\angle(\cdot)$ denotes the angle of a complex number, $\alpha\in(0,1]$ is an amplitude companding constant, and $\beta$ is the scaling constant to normalize the final amplitudes roughly within $[0,1]$, and the corresponding demodulation 
\begin{align}
\boldsymbol{x}_{c}=\beta^{-1}|\boldsymbol{x}_a|^{\frac{1}{\alpha}} e^{i\angle(\boldsymbol{x}_a)}
\label{eq:dm}
\end{align}
is applied to the enhanced complex spectra before the inverse STFT.

\section{Experiments}
\label{sec:typestyle}
\vspace{-\baselineskip}
\subsection{Experimental Setting}
This paper focuses on speech modeling because of the majority of human communication. Specifically, the codecs were evaluated on the full-band 48~kHz VCTK~\cite{vctk2017}-derived Valentini~\cite{noisyvctk} dataset consisting of 84 gender-balanced English speakers for training and two speakers (female p257 and male p232) for testing. Each speaker has around 400 utterances, and the utterance lengths vary from 1--16s.

The symmetric AudioDec (symAD) and AudioDec~\cite{audiodec} codecs were adopted as the baselines. The proposed ScoreDec and symAD share the pre-trained AutoEncoder (AE) as shown in Fig.~\ref{fig:scoredec} while the symAD decoder was further trained with the GAN training. ScoreDec in contrast uses the score-based post-filter (SPF). The vocoder-based AudioDec codec, which replaced the symAD decoder with a powerful HiFi-GAN vocoder, was also included for fair comparisons. In addition to the mel loss, the wrapped angle loss from Eq.~\ref{eq:phase_wrap} ($L_{angle}$), the multi-resolution mel loss ($L_{mm}$)~\cite{pwg}, and the L1 waveform loss ($L_{wav}$) were applied to the symAD and AudioDec codecs to show the difficulties of phase modeling. The model architectures and hyperparameters of the pre-trained AE and HiFi-GAN follow~\cite{audiodec}\footnote{\label{repo}\url{https://github.com/facebookresearch/AudioDec}} but the compression rate and the number of the 10-bit codebooks were increased to 320 and 16 to achieve a bitrate of 24~kbps. The AE was trained with only the metric losses for the first 500k~iterations, and the decoder/vocoder was further trained with the discriminators for another 500k~iterations.

Opus~\cite{opus} with 24~kbps, which maintains only the most audio information under 20~kHz, was also adopted to show the effectiveness of the proposed SPF for different codecs. The SPFs of ScoreDec and Opus followed the SGMSE\footnote{\url{https://github.com/sp-uhh/sgmse}} architecture. The complex spectra were extracted using 510 FFT size and 320 hop size. The real and imaginary spectra were treated as two separate channels in the models. The stochastic parameters were set to  $\sigma_{\text{min}}=0.05$, $\sigma_{\text{max}}=0.5$, and $\gamma=1.5$. The diffusion time step was within $[0.03, 1]$. The modulation parameters were set to $\alpha=0.5$ and $\beta=0.15$. The batch size was 8 and the batch length was 256 frames. Both models were trained for 161 epochs with a learning rate of 10\textsuperscript{-4}. The predictor-corrector (PC) sampler ~\cite{sgm} with one corrector step was adopted for inference. The step size of the annealed Langevin dynamic in the corrector was 0.5. We run the sequential inference of the score-based model for 30 steps. More details can be found in~\cite{sgmse2}.

\begin{table}[t]
\caption{Objective evaluations of 48~kHz codecs w/ 24~kbps}
\label{tb:objective}
\fontsize{7pt}{9.6pt}
\selectfont
{%
\begin{tabularx}{1.0\columnwidth}{@{}p{2.2cm}YYYY@{}}
\toprule
& Wav($\times$10\textsuperscript{-3})$\downarrow$ & SI-SDR$\uparrow$ &STOI$\uparrow$ &PESQ$\uparrow$ \\ \midrule
symAD       &8.5   &-17.38   & 0.91   & 3.12   \\
symAD w/ $L_{angle}$ &2.6   &0.70   & 0.90   & 2.60   \\
symAD w/ $L_{mm}$    &2.9   &0.29   & 0.91   & 2.51   \\
symAD w/ $L_{wav}$   &1.3   &5.00   & 0.91   & 2.57   \\

AudioDec          &3.1   &-0.50  & 0.91   & 2.67   \\
AudioDec w/ $L_{angle}$    &2.4   &1.34   & 0.90   & 2.58   \\
AudioDec w/ $L_{mm}$       &2.6   &0.76   & 0.92   & 2.53   \\
AudioDec w/ $L_{wav}$      &1.2   &5.19   & 0.92   & 2.60   \\ 
\textbf{ScoreDec (ours)}          &\textbf{0.7}   &\textbf{8.17}   & \textbf{0.97}   & \textbf{3.68}   \\
\midrule
Opus        &10.0   &-20.62    & 0.89   & 4.21   \\ 
\textbf{Opus\_SPF (ours)}    &\textbf{0.2}   &\textbf{16.20}   & \textbf{0.98}   & \textbf{4.29}   \\
\bottomrule
\end{tabularx}%
}
\end{table}

\begin{figure}[t]
\centering
\centerline{\includegraphics[width=1\columnwidth]{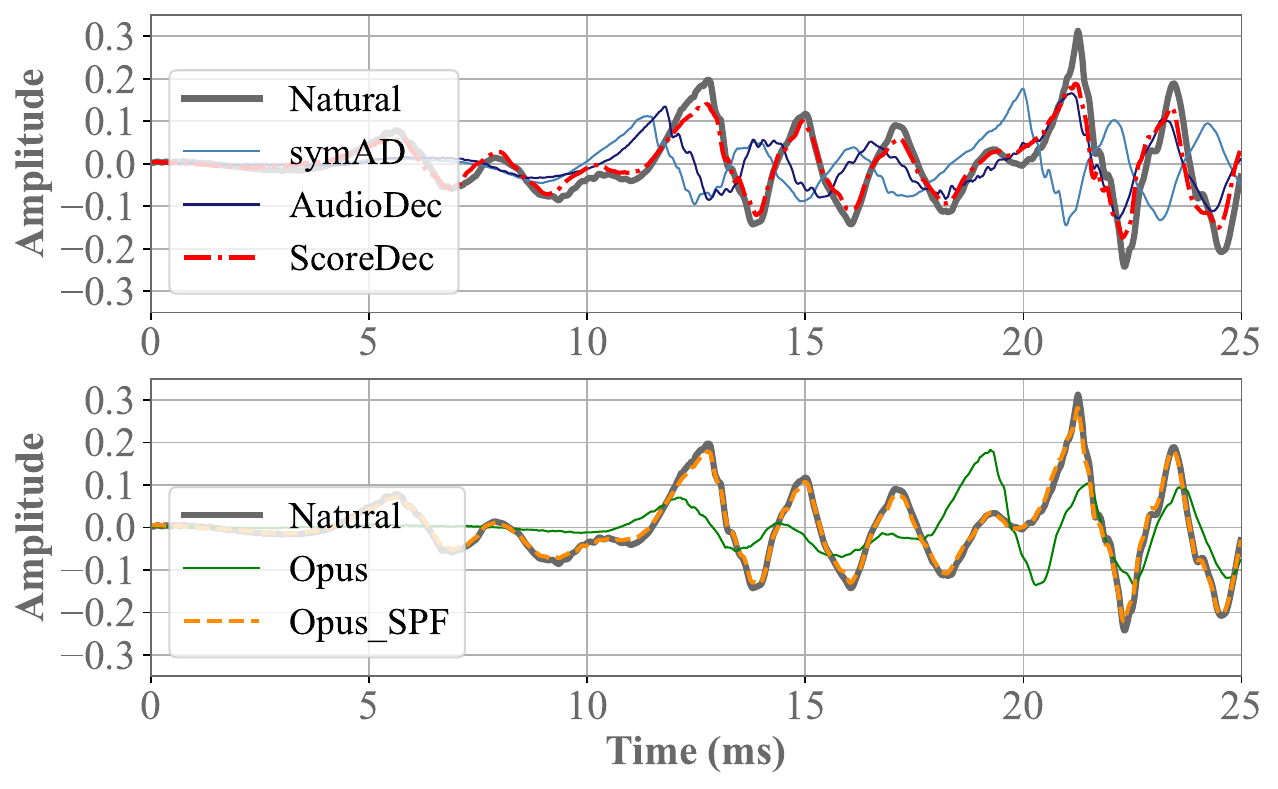}}
\caption{Waveform comparison of the existing neural audio codec AudioDec to ScoreDec (top) and of Opus with and without our proposed score-based diffusion post-filter (bottom). ScoreDec and Opus\_SDF preserve the original phase well.}
\label{fig:waveform}
\end{figure}

\begin{table}[t]
\caption{Mean Opinion Scores of 48~kHz codecs w/ 24~kbps}
\label{tb:subjective}
\fontsize{7pt}{9.6pt}
\selectfont
{%
\begin{tabularx}{\columnwidth}{@{}YYYYYY@{}}
\toprule
Natural & symAD & AudioDec  & \textbf{ScoreDec} & Opus & \textbf{Opus\_SPF}\\ \midrule

4.14$\pm$.08 & 3.86$\pm$.09 & 3.79$\pm$.10   & \textbf{4.16}$\pm$.08   & 3.88$\pm$.09   & \textbf{4.16}$\pm$.08 \\
\bottomrule
\end{tabularx}%
}
\end{table}

\subsection{Objective Evaluation}
To evaluate the phase-preserving capability via the waveform similarity, we report waveform mean-square-error (Wav) and scale-invariant source-to-distortion ratio (SI-SDR)~\cite{sisdr} in dB. To evaluate the speech intelligibility and quality, wide-band STOI~\cite{stoi} and PESQ~\cite{pesq} were applied to the downsampled 16~kHz speech. The results are the averages of all testing utterances. As shown in Table~\ref{tb:objective}, the proposed ScoreDec and Opus\_SPF respectively outperform their base codecs in all measurements, especially the waveform similarity. The much higher SI-SDR and greatly reduced Wav loss (10\textsuperscript{-3}$\rightarrow$10\textsuperscript{-4}) demonstrate the significant phase-preserving improvement of the proposed models while achieving even higher speech intelligibility and quality. On the other hand, the markedly worse waveform similarity of models adopting the explicit or implicit phase modeling losses also show the difficulties of direct phase modeling. That is, although these losses can improve the waveform similarity, the waveform similarity gap is still significant, and the speech quality also slightly degrades.  

To demonstrate the outstanding phase preservation in ScoreDec, we plot a waveform example in Fig.~\ref{fig:waveform}. The results indicates that although the symAD-generated waveform has a similar magnitude spectrum to the natural one as shown in Fig.~\ref{fig:mag_phase}, the phase is very different from the natural one. Although AudioDec achieves a better waveform similarity, the phase differences are still significant. Even the DSP-based Opus codec has the same problem. However, the ScoreDec- and Opus\_SPF-generated waveforms well align with the natural waveform, showing that the original phase is well preserved.

On the other hand, we also evaluated the inference speed using an NVIDIA A100 80GB SXM GPU. ScoreDec with a real-time factor (RTF) of 1.707 is significantly slower than symAD with a 0.019 RTF and AudioDec with a 0.022 RTF as expected. The non-causal architecture and iterative inference hider ScoreDec from streaming applications, and we leave fast inference as an important future work. 

\subsection{Subjective Evaluation}
Mean opinion score (MOS) tests were conducted to evaluate the perceptual quality. We randomly selected 15 utterances of each test speaker to form the testing set including natural and codec-generated speech. Fifteen participants took the tests in a quiet environment with headphones. The participants are either native speakers or audio researchers. The score range is 1 (very unnatural)--5 (very natural), and the average score with a 95\% confidence interval of each system is shown in Table~\ref{tb:subjective}. The results show that an E2E neural codec such as symmetric AudioDec achieves only similar speech quality to the DSP-based Opus codec when adopting the normal Opus operation bitrate, 24~kbps. Even increasing the model capacity by using a powerful HiFi-GAN in AudioDec, the quality improvements of the generated speech are saturated. However, the proposed SPF can further improve the speech quality to a human-level naturalness. Moreover, SPF generalizes well to both neural- and DSP-based codecs, since both ScoreDec and Opus\_SPF achieve human-level naturalness. More details can be found on our demo page\footnote{\url{https://bigpon.github.io/ScoreDec\_demo/}}.

\section{Conclusion}
\label{sec:conclusion}
\vspace{-\baselineskip}
In this paper, we demonstrate that score-based diffusion post-filtering improves existing neural- and DSP-based audio codecs to human-level naturalness in speech modeling, and adversarial training is not required since high-fidelity results can be achieved exclusively using metric and score-matching losses. The proposed ScoreDec preserves phase information well and therefore renders itself suitable not only for mono audio codecs but also for multichannel codecs. For future work, the effectiveness of ScoreDec for modeling general spatial audio signals beyond speech should be evaluated. A streamable ScoreDec with a causal architecture and fast inference is also an important topic to extend the ScoreDec capability from the limited non-streaming applications.




\vfill\pagebreak

\bibliographystyle{IEEEbib}
\bibliography{refs}

\begin{thebibliography}{10}

\bibitem{mpeg4}
Tilman Liebchen and Yuriy~A Reznik,
\newblock ``{MPEG-4 ALS}: An emerging standard for lossless audio coding,''
\newblock in {\em Proc. DCC}, 2004, pp. 439--448.

\bibitem{flac}
J.~Coalson,
\newblock {\em Free {L}ossless {A}udio {C}odec}, Accessed: 2000.

\bibitem{opus}
J.-M. Valin, G.~Maxwell, T.~B. Terriberry, and K.~Vos,
\newblock ``High-quality, low-delay music coding in the opus codec,''
\newblock in {\em AESC 135}, 2013.

\bibitem{amrwb}
B.~Bessette et~al.,
\newblock ``The adaptive multirate wideband speech codec {(AMR-WB)},''
\newblock {\em IEEE TSAP}, vol. 10, no. 8, pp. 620--636, 2002.

\bibitem{evs}
M.~Dietz et~al.,
\newblock ``Overview of the {EVS} codec architecture,''
\newblock in {\em Proc. ICASSP}, 2015, pp. 5698--5702.

\bibitem{aecodec2018}
S.~Kankanahalli,
\newblock ``End-to-end optimized speech coding with deep neural networks,''
\newblock in {\em Proc. ICASSP}, 2018, pp. 2521--2525.

\bibitem{vqvae2019}
C.~G{\^a}rbacea and other,
\newblock ``Low bit-rate speech coding with vq-vae and a wavenet decoder,''
\newblock in {\em Proc. ICASSP}, 2019, pp. 735--739.

\bibitem{cmrl}
K.~Zhen, J.~Sung, M.~S. Lee, S.~Beack, and M.~Kim,
\newblock ``Cascaded cross-module residual learning towards lightweight
  end-to-end speech coding,''
\newblock in {\em Proc. Interspeech}, 2019, pp. 3396--3400.

\bibitem{soundstream}
N.~Zeghidour, A.~Luebs, A.~Omran, J.~Skoglund, and M.~Tagliasacchi,
\newblock ``{SoundStream}: An end-to-end neural audio codec,''
\newblock {\em IEEE/ACM TASLP}, vol. 30, pp. 495--507, 2021.

\bibitem{encodec}
A.~D{\'e}fossez, J.~Copet, G.~Synnaeve, and Y.~Adi,
\newblock ``High fidelity neural audio compression,''
\newblock {\em arXiv preprint arXiv:2210.13438}, 2022.

\bibitem{audiodec}
Y.-C. Wu, I.~D. Gebru, D.~Markovi{\'c}, and A.~Richard,
\newblock ``Audio{D}ec: An open-source streaming high-fidelity neural audio
  codec,''
\newblock in {\em Proc. ICASSP}, 2023.

\bibitem{rvq}
A.~Vasuki and P.T. Vanathi,
\newblock ``A review of vector quantization techniques,''
\newblock {\em IEEE Potentials}, vol. 25, no. 4, pp. 39--47, 2006.

\bibitem{gan}
I.~Goodfellow, J.~Pouget-Abadie, M.~Mirza, B.~Xu, D.~Warde-Farley, S.~Ozair,
  A.~Courville, and Y.~Bengio,
\newblock ``Generative adversarial nets,''
\newblock in {\em Proc. Proc. NeurIPS}, 2014, pp. 2672--2680.

\bibitem{ddpm}
J.~Ho, A.~Jain, and P.~Abbeel,
\newblock ``Denoising diffusion probabilistic models,''
\newblock in {\em Proc. NeurIPS}, 2020.

\bibitem{sgm}
Y.~Song, J.~Sohl-Dickstein, D.~P. Kingma, A.~Kumar, S.~Ermon, and B.~Poole,
\newblock ``Score-based generative modeling through stochastic differential
  equations,''
\newblock in {\em Proc. ICLR}, 2021.

\bibitem{diffwave}
Z.~Kong, W.~Ping, J.~Huang, K.~Zhao, and B.~Catanzaro,
\newblock ``Diff{W}ave: A versatile diffusion model for audio synthesis,''
\newblock in {\em Proc. ICLR}, 2021.

\bibitem{wavegrad}
N.~Chen, Y.~Zhang, H.~Zen, R.~J. Weiss, M.~Norouzi, and W.~Chan,
\newblock ``Wave{G}rad: Estimating gradients for waveform generation,''
\newblock in {\em Proc. ICLR}, 2021.

\bibitem{sgmse1}
S.~Welker, J.~Richter, and T.~Gerkmann,
\newblock ``Speech enhancement with score-based generative models in the
  complex {STFT} domain,''
\newblock in {\em Proc. Interspeech}, 2022, pp. 2928--2932.

\bibitem{sgmse2}
J.~Richter, S.~Welker, J.-M. Lemercier, B.~Lay, and T.~Gerkmann,
\newblock ``Speech enhancement and dereverberation with diffusion-based
  generative models,''
\newblock {\em IEEE/ACM TASLP}, vol. 31, pp. 2351--2364, 2023.

\bibitem{cdse}
Y.-J. Lu, Z.-Q. Wang, S.~Watanabe, A.~Richard, C.~Yu, and Y.~Tsao,
\newblock ``Conditional diffusion probabilistic model for speech enhancement,''
\newblock in {\em Proc. ICASSP}, 2022, pp. 7402--7406.

\bibitem{melgan}
K.~Kumar et~al.,
\newblock ``{MelGAN}: generative adversarial networks for conditional waveform
  synthesis,''
\newblock in {\em Proc. NeurIPS}, 2019.

\bibitem{hifigan}
J.~Kong, J.~Kim, and J.~Bae,
\newblock ``{HiFi-GAN}: Generative adversarial networks for efficient and high
  fidelity speech synthesis,''
\newblock in {\em Proc. NeurIPS}, 2020.

\bibitem{ouve}
G.~E. Uhlenbeck and L.~S. Ornstein,
\newblock ``On the theory of the brownian motion,''
\newblock {\em Physical review}, vol. 36, no. 5, pp. 823, 1930.

\bibitem{rdem}
B.~D. Anderson,
\newblock ``Reverse-time diffusion equation models,''
\newblock {\em Stochastic Processes and their Applications}, vol. 12, no. 3,
  pp. 313--326, 1982.

\bibitem{score_match}
A.~Hyv{\"a}rinen and P.~Dayan,
\newblock ``Estimation of non-normalized statistical models by score
  matching.,''
\newblock {\em Journal of Machine Learning Research}, vol. 6, no. 4, 2005.

\bibitem{nspp}
Y.~Ai and Z.-H. Ling,
\newblock ``Neural speech phase prediction based on parallel estimation
  architecture and anti-wrapping losses,''
\newblock in {\em Proc. ICASSP}, 2023.

\bibitem{vctk2017}
C.~Veaux, J.~Yamagishi, and K.~MacDonald,
\newblock ``{CSTR VCTK} corpus: English multi-speaker corpus for {CSTR} voice
  cloning toolkit,''
\newblock {\em University of Edinburgh. CSTR}, 2017.

\bibitem{noisyvctk}
C.~Valentini-Botinhao,
\newblock ``Noisy speech database for training speech enhancement algorithms
  and {TTS} models,''
\newblock {\em University of Edinburgh. CSTR}, 2017.

\bibitem{pwg}
R.~Yamamoto, E.~Song, and J.-M. Kim,
\newblock ``Parallel {W}ave{GAN}: A fast waveform generation model based on
  generative adversarial networks with multi-resolution spectrogram,''
\newblock in {\em Proc. ICASSP}, 2020, pp. 6199--6203.

\bibitem{sisdr}
J.~L. Roux, S.~Wisdom, H.~Erdogan, and J.~R. Hershey,
\newblock ``Sdr--half-baked or well done?,''
\newblock in {\em Proc. ICASSP}, 2019, pp. 626--630.

\bibitem{stoi}
C.~H. Taal, R.~C. Hendriks, R.~Heusdens, and J.~Jensen,
\newblock ``An algorithm for intelligibility prediction of time--frequency
  weighted noisy speech,''
\newblock {\em IEEE/ACM TASLP}, vol. 19, no. 7, pp. 2125--2136, 2011.

\bibitem{pesq}
A.~Rix, J.~Beerends, M.~Hollier, and A.~Hekstra,
\newblock ``Perceptual evaluation of speech quality ({PESQ})-a new method for
  speech quality assessment of telephone networks and codecs,''
\newblock in {\em Proc. ICASSP}, 2001, vol.~2, pp. 749--752.

\end{thebibliography}

\end{document}